\documentclass[aps,prd,nofootinbib]{revtex4}
\usepackage[utf8]{inputenc}
\usepackage[english]{babel}
\usepackage[T1]{fontenc}

\usepackage{calligra}
\usepackage{dutchcal} 

\usepackage{amsmath}
\usepackage{amsfonts}
\usepackage{amssymb}
\usepackage{graphicx}

\usepackage{lmodern}

\usepackage{hyperref}
\hypersetup{colorlinks, 
citecolor=blue,
linkcolor=red,
urlcolor=black}

\usepackage[normalem]{ulem}

\usepackage{booktabs}

\begin{document}

\title{Quantum Brownian motion induced by a scalar field in Einstein's universe}
\author{E. J. B. Ferreira}
\email{ejbf@academico.ufpb.br}
\author{H. F. Santana Mota}
\email{hmota@fisica.ufpb.br}
\affiliation{Departamento de Física, Universidade Federal da Paraíba, Caixa Postal 5008, João Pessoa, Paraíba, Brazil}

%\date{\today}

\begin{abstract}
The Brownian motion of a point particle induced by quantum vacuum fluctuations of a massless real scalar field in Einstein's universe is studied. By assuming the small displacement condition, the dispersion in the momentum and position of a point particle coupled to the massless scalar field are obtained. As a consequence of the homogeneity and isotropy properties of the Einstein's Universe, we find that all components of these physical observables are identical. We also examine divergent behaviors associated with the physical momentum and position dispersions, which we attribute to the I\!R$^{1}\times$S$^{3}$ compact topology of Einstein's universe. Finally, based on the small displacement condition assumed, we analyze the limit of validity of our investigation. 
\end{abstract}

%\pacs{87.01-a}

\maketitle

\section{Introduction}

The stochastic motion that a small point particle can undergo as a consequense of its interaction with quantum fields has been increasingly studied in recent decades considering the most diverse scenarios and aspects \cite{gour1999will,yu2004brownian,yu2004vacuum,yu2006brownian,seriu2008switching,seriu2009smearing, de2014quantum,de2016probing,camargo2018vacuum,de2019remarks,camargo2019vacuum,Camargo:2020fxp, Ferreira:2023uxs,bessa2009brownian,bessa2017quantum,mota2020induced,anacleto2021stochastic,ferreira2022quantum}. The fluctuations associated with quantum fields (by virtue of their vacuum state, for instance) may produce effects on the motion of classical test particles. Such a phenomenon, of quantum origin, is completely aleatory and induce small random deviations in the classical paths of the particles. Mathematicaly, these effects can be analyzed through the calculation of the dispersion associated with physical observables characterizing the particle as, for example, velocity (or momentum) and position. The random quantum motion arising in this framework resembles, in some aspects, the classical Brownian motion problem of a particle suspended in a fluid. In view of the similarities, it is common to use the terminology {\it induced quantum Brownian motion} (IQBM), which is the one to be adopted here.

In general, IQBM investigations consider the Minkowski spacetime, thus, ignoring gravity effects. In this sense, the nonzero velocity and position dispersions of the classical particle steam from different conditions applied on the quantum field \cite{gour1999will,yu2004brownian,yu2004vacuum,yu2006brownian,seriu2008switching,seriu2009smearing, de2014quantum,de2016probing,camargo2018vacuum,de2019remarks,camargo2019vacuum,Camargo:2020fxp, Ferreira:2023uxs}. On the other hand, the study of the IQBM in curved spacetime automatically adds extra difficulties, since gravity effects contributions must be taken into consideration, which leads to more complicated equations of motion for both the field and the particle. In conformally flat spacetimes, as the one described by the Friedmann-Robertson-Walker (FRW) line element, the IQBM has been considered in Refs. \cite{bessa2017quantum, mota2020induced} for scalar fields. Conformally flat spacetimes are of particular interest since the symmetries involved allow us to solve the problem in a fashionable way. In addition, the effects of spacetime topology on the motion of point particles coupled to a quantized electromagnetic field has also been investigated by making use of the conformally flat spacetime symmetry in Refs. \cite{bessa2009brownian, Bessa:2019aar, Lemos:2021jzy}. In this paper, we investigate the IQBM of a point particle coupled to a massless quantum scalar field in a spacetime whose geometry is described by the Einstein's universe, a curved spacetime with positive constant curvature. This is obtained from the FRW spacetime, with closed spatial section, by considering a constant scale factor. Note that this spacetime is not conformally flat. In Ref. \cite{Hu:1993qa} the authors considered a Brownian particle coupled to a bath of time dependent quadratic oscillators.

The contributions of quantum vacuum fluctuations due to the Einstein's universe have already been extensively investigated in the context of Casimir effect \cite{Kennedy:1980kc,Ozcan:2006jn,ford1975quantum,ford1976quantum,Mota:2015ppk,Mota:2022qpf, Bezerra:2021qnw}. In contrast, our investigation consider how geometrical aspects associated with the closed curvature of the Einstein's universe contribute to produce IQBM. Note that the geometry of the Einstein's universe has also been discussed in the cosmological scenario \cite{ellis2003emergent}. Moreover, a recent experiment based on a Bose-Einstein condensate has been proposed in order to simulate an expanding spacetime geometry (like the FRW model of cosmology), considering negative and positive curvatures as well \cite{benini2023ultracold,weinfurtner2022superfluid,Viermann:2022wgw}. Therefore, motivated by the several scenarios where this geometry is considered, our study has a fundamental importance of exploring the IQBM phenomenon in the curved spacetime described by the Einstein's universe, a investigation that is conducted for the first time in the present paper, to the best of our knowledge. 

Regarding the structure of this work, in Section \ref{Sec2} we briefly present both the spacetime geometry in which we carry out our investigation and the solution of the Klein-Gordon equation, also obtaining the positive frequency Wightman function. In Section \ref{Sec3} we establish the expressions referring to the dispersion of the momentum and position of the particle and study their behaviors. Finally, we present our conclusions summarizing the main points and results.

\section{Curved space-time, Normalizated solutions and Wightman function}\label{Sec2}
%
%Nessa seção nós vamos estabelecer os elementos necessários para estudar o MBI de uma partícula acoplada com um campo escalar sem massa no universo de Einstein. Um elemento crucial em nossos cálculos é a função de Wightman de frequência positiva. Contudo, para obter essa quantidade primeiro é necessário encontrar as soluções normalizadas (modos) da equação de Klein-Gordon no Universo de Einstein e construir o operador de campo. Em seguida esse processo é descrito em detalhes. Adiantamos que para obter as soluções da equação Klein-Gordon e a função de Wightman nós temos nos baseado nas Refs. \cite{Ozcan:2001cr,Ozcan:2006jn}.

In this section we will establish the necessary elements to study the IQBM of a point particle coupled to a massless quantum scalar field in Einstein's universe. A crucial element in our calculations is the positive frequency Wightman function. To btain this quantity we first need to find the normalized solutions (modes) of the Klein-Gordon equation in Einstein's universe and construct the field operator. In the following, this process is described in detail. In order to obtain the Klein-Gordon solution and Wightman function we based our analysis on Refs. \cite{Ozcan:2001cr,Ozcan:2006jn}.

\subsection{Curved space-time background geometry: Einstein's universe.}

The Friedmann-Robertson-Walker (FRW) spacetime describes the standard geometric structure of relativistic cosmology, which satisfies constraints based on observational facts, for example, the expansion of the Universe, the homogeneity and isotropy of the large-scale Universe, etc \cite{adler2021general}. As we know, in the metric characterizing the spacetime two elements of fundamental importance are the curvature parameter $k$ and the scale factor $a$. The time dependent function $a(t)$ is real and gives the form of the accelerated expansion of the Universe. The $k$ parameter is a constante and can take on three specific values, namely, $k=(-1,0,+1)$, which specify distinct geometry and topology for the spacetime, but with all the cases being equally homogeneous and isotropic \cite{adler2021general,schutz2009first}. Here, we are interested in the $k=+1$ case of the FRW metric, which defines the Einstein's static universe, whose mathematical structure of the corresponding line element is given by \cite{schutz2009first,d1992introducing,islan2004}
\begin{eqnarray}\label{metricEinsteinUniverse}
ds^{2} = dt^{2}-a^{2}_{0}\left\{d\chi^{2}+\sin^{2}(\chi)\left[d\theta^{2}+\sin^{2}(\theta)d\phi^{2}\right]\right\},
\end{eqnarray}
where $a_{0}=a(t=t_{0})$ represents a constante scale factor defined by a hypersurface of constante time $t=t_{0}$, which is identified as the radius of the Einstein's universe. Hence, Eq. \eqref{metricEinsteinUniverse} defines the line element of the Einstein's universe and describes a closed and static spacetime with radius $a_{0}$, $0\leq\chi\leq\pi$, $0\leq\theta\leq\pi$ and $0\leq\phi\leq 2\pi$. As we will see later in this paper, since the Einstein's universe has a completely closed geometry, the modes of a scalar field in this spacetime will be subject to confinement-like effects, in other words, the quantum modes will naturally be discretized. Throughout this paper we will use natural units, so that $c=\hbar=1$.

%onde $a_{0}=a(t=t_{0})$ representa um fator de escala constante definido por uma hipersuperfície de tempo constante $t=t_{0}$, que é identificado como o raio do universo de Einstein. A eq. \eqref{metricEinsteinUniverse} define o elemento de linha do universo de Einstein, o qual descreve um espaço-tempo fechado. Como veremos posteriormente nesse trabalho, uma vez que o universo de Einstein possui uma geometria completamente fechada os modos de um campo escalar nesse espaço-tempo estarão sujeitos a um efeito similar ao de um confinamento, isto é, naturalmente todos os momentos serão discretizados.

\subsection{Modes}

As we are interested in studying the IQBM as a consequence of quantum vacuum fluctuations of a massless real scalar field in the Einstein's universe described by the line element \eqref{metricEinsteinUniverse}, we need now to solve the Klein-Gordon equation
\begin{eqnarray}\label{eqKG}
(\Box + m_{F}^{2} + \xi R)\psi(x) = 0,
\end{eqnarray}
where $\Box \psi(x)$ is the D’Alembertian differential operator in curved spacetime \cite{birrell1984quantum}.
%
%\begin{eqnarray}\label{sec3eq00}
%\Box \psi(x) = \frac{1}{\sqrt{-g}}\partial_{\mu}[\sqrt{-g}g^{\mu\nu}\partial_{\nu}\psi(x)],
%\end{eqnarray}
%
%with $g = \textrm{Det}(g_{\mu\nu})$ and $g_{\mu\nu}$ obtained from Eq. \eqref{metricEinsteinUniverse}. 
The parameter $m_{F}^{2}$ is the field mass and $\xi$ is the coupling constant of the scalar field $\psi(x)$ to gravity. In the cases $\xi=0$ and $\xi\neq 0 $ we have, respectively, a minimally and non-minimally coupling to gravity. On the other hand, when $\xi=(n-2)/4(n-1)$ we have the conformally coupled case, where $n$ is the spacetime dimension number \cite{birrell1984quantum}. Here $n=4$, so that the constant $\xi=1/6$.
The object $R(x)$ is the Ricci scalar, which can be obtained in terms of the Ricci tensor $R_{\mu\nu}(x)$ through the expression $R=g^{\mu\nu}R_{\mu\nu}$ \cite{adler2021general,schutz2009first}.

%, with \cite{adler2021general}
%
%\begin{eqnarray}\label{RicciTensor}
%R_{\mu\nu}(x) = \Gamma^{\beta}_{\beta\nu,\mu}-\Gamma^{\beta}_{\mu\nu,\beta}+\Gamma^{\beta}_{\alpha\mu}\Gamma^{\alpha}_{\beta\nu}-\Gamma^{\beta}_{\alpha\beta}\Gamma^{\alpha}_{\mu\nu},
%\end{eqnarray}
%
%where the objects $\Gamma^{\alpha}_{\mu\nu}$ are the Christoffel symbols and can be obtained from the metric tensor as follows
%
%\begin{eqnarray}\label{sec4eq00b}
%\Gamma^{\alpha}_{\mu\nu} = \frac{1}{2}g^{\alpha\beta}\left( g_{\beta\mu,\nu}+g_{\beta\nu,\mu}-g_{\mu\nu,\beta}\right).
%\end{eqnarray}
%

The first step in solving Eq. \eqref{eqKG} is to assume separable solutions, that is, consider that scalar field is decomposed in independent solutions for each variables:
\begin{eqnarray}\label{sec3eq01}
\psi(t,\chi,\theta,\phi)=T(t)\mathcal{R}(\chi)\Theta(\theta)F(\phi).
\end{eqnarray}
Thus, substituting \eqref{sec3eq01} in Eq. \eqref{eqKG}, we easily obtain that the solution for the temporal part of $\psi$ is given by 
\begin{eqnarray}\label{sec3eq02a}
T(t) = T_{0}e^{-i\omega t},
\end{eqnarray}
where $T_{0}$ is a constant and we define $\omega^{2}= (k/a_{0})^{2}+M^{2}$, with $M^{2}= m_{F}^{2}+\xi R$. Mathematically, the parameters $\omega$ and $k$ are separation constants that arising from the ansatz \eqref{sec3eq01}, but, as we will see, they are related to the frequencies and quantum numbers of the modes, respectively. 

Similarly, after some computation,  we find that the angular parts $\theta$ and $\phi$ correspond to the usual spherical harmonics $Y_{\ell}^{m}(\theta,\phi)$, namely,
\begin{eqnarray}\label{sec3eq02b}
\Theta(\theta)F(\phi)\equiv Y_{\ell}^{m}(\theta ,\phi) = (-1)^{m}\sqrt{\dfrac{(2\ell+1)}{4\pi}\frac{(\ell - m)!}{(\ell + m!)}}P_{\ell}^{m}(\cos(\theta))e^{im\phi},
\end{eqnarray}
where $P_{\ell}^{m}$ are the associate Legendre functions \cite{arfken2005mathematical}, with $0\leq\phi\leq 2\pi$, $0\leq\theta\leq\pi$, $\ell=\{0,1,2,\ldots\}$ and $-\ell\leq m\leq\ell$.

Finally, in order to solve the equation in the angular variable $\chi$ \cite{ford1975quantum}, that is,
\begin{eqnarray}\nonumber
\dfrac{\partial}{\partial\chi}\left[\sin^{2}(\chi)\dfrac{\partial\mathcal{R}}{\partial\chi}\right]+ [k^{2}\sin^{2}(\chi)-\ell(\ell +1)]\mathcal{R}=0,
\end{eqnarray}
we assume that $\mathcal{R}(\chi)=\sin^{\ell}(\chi)f(\chi)$ and perform the chance of variable $z=\cos(\chi)$, so that we obtain
\begin{eqnarray}\label{sec3eqgegenbauer1}
(1-z^{2})f''-[2(\ell +1)+1]zf'+[k^{2}-\ell(\ell +2)]f=0,
\end{eqnarray}
where the prime symbol means derivative with respect to $z$. Note that in this new variable we have the range correspondence $\chi=[0,\pi]$ to $z=[-1,1]$.

Observing Eq. \eqref{sec3eqgegenbauer1} we note that its structure is similar to the differential equation
\begin{eqnarray}\label{sec3eqgegenbauer2}
(1-z^{2})\dfrac{d^{2}g(z)}{dz^{2}} - (2\alpha +1)\dfrac{dg(z)}{dz}+m(2\alpha +m)g(z)=0,
\end{eqnarray}
whose solutions are the functions $C^{\alpha}_{m}(z)$, known as Gegenbauer polynomials or ultraspherical polynomials \cite{abramowitz1970handbook}, where $\alpha$ is an arbitrary number and $m$ a natural number that corresponds to the order of the polynomial.
Legendre polynomials are a particular case of the Gegenbaur polynomials for $\alpha = 1/2$, namely, $C_{m}^{(1/2)}(z) = P_{m}(z)$ \cite{gradshtein2007}. 
So, by making the correspondence $m\rightarrow n-\ell$ and $\alpha\rightarrow\ell +1$ in Eq. \eqref{sec3eqgegenbauer1} and identifying $k^{2}\equiv n(n+2)$ in Eq. \eqref{sec3eqgegenbauer2}, we obtain that $f(z)= C_{n-\ell}^{\ell +1}(z)$ and consequently \cite{ford1975quantum,ford1976quantum,Ozcan:2001cr}
\begin{eqnarray}\label{sec3eq02c}
\mathcal{R}(\chi) = \sin^{\ell}(\chi)C_{n-\ell}^{\ell+1}(\cos(\chi)),
\end{eqnarray}
where $n=0,1,2,3,...$\;.

In view of the results \eqref{sec3eq02a}, \eqref{sec3eq02b} and \eqref{sec3eq02c}, from Eq. \eqref{sec3eq01}, we obtain that
\begin{eqnarray}\label{sec3modes}
\psi_{\sigma}(t,\chi,\theta,\phi) = \textrm{N}\sin^{\ell}(\chi)C_{n-\ell}^{\ell +1}(\cos\chi)Y_{\ell}^{m}(\theta,\phi)e^{-i\omega_{n} t}
\end{eqnarray}
are the mode solutions that satisfy Eq. \eqref{eqKG}, 
\begin{eqnarray}\label{sec3eigenfreq}
\omega_{n} = \left[ \dfrac{n(n+2)}{a_{0}^{2}} + M^{2} \right]^{1/2},
\end{eqnarray}
are the eigenfrequencies and $\sigma = (n,\ell,m)$ stands for the set of field's modes. The constant N can be obtained from the normalization condition \cite{birrell1984quantum}
\begin{eqnarray}\label{sec3normalization}
-i\int dx^{3}\sqrt{-g}[\psi_{\sigma}(\partial_{t}\psi^{*}_{\sigma'})-(\partial_{t}\psi_{\sigma})\psi^{*}_{\sigma'}]=\delta_{\sigma\sigma'},
\end{eqnarray}
where $\delta_{\sigma\sigma'}$ stands for Kronecker delta in the case the mode is discrete and for Dirac delta in the case the mode is continuous. 

From Eqs. \eqref{sec3modes} and \eqref{sec3normalization} we obtain
\begin{eqnarray}\label{sec3freemodessol1}
\psi_{\sigma}(t,\chi,\theta,\phi) = \textrm{N}_{n\ell}\sin^{\ell}(\chi)C_{n-\ell}^{\ell +1}(\cos\chi)Y_{\ell}^{m}(\theta,\phi)e^{-i\omega_{n} t},
\end{eqnarray}
where
\begin{eqnarray}\label{sec3freemodessol2}
\textrm{N}_{n\ell}=\left\{ \dfrac{2^{2\ell}(n+1)(n-\ell)![\Gamma(\ell +1)]^{2}}{\pi a_{0}^{3}\omega_{n}\Gamma(\ell +n +2)}\right\}^{1/2}.
\end{eqnarray}	
Note that the eigenfrequencies $\omega_{n}$ are defined in Eq. \eqref{sec3eigenfreq}. To obtain the above equation, from Eq. \eqref{metricEinsteinUniverse} we note that $\sqrt{-g}=a_{0}^{3}\sin^{2}(\chi)\sin(\theta)$. Furthermore, we have used orthogonality relations for the spherical harmonics \cite{arfken2005mathematical}, 
%
%\begin{eqnarray}\label{prop_sphe_har_ort}
%\int_{0}^{2\pi}d\phi\int_{0}^{\pi}d\theta\sin(\theta)[Y_{\ell}^{m}(\theta,\phi)]^{*}Y_{\ell'}^{m'}(\theta,\phi) = \delta_{\ell\ell'}\delta_{mm'},
%\end{eqnarray}
%
and for the Gegenbauer polynomials \cite{gradshtein2007, abramowitz1970handbook}.
%
%\begin{eqnarray}\label{prop_gegenb_ort}
%\int_{-1}^{1}dz(1-z^2)^{\lambda -1/2}C^{\lambda}_{j}(z)C^{\lambda}_{k}(z) = \left\{
%	\begin{array}{ll}
%		&0\  \textrm{if} \ j\neq k, \\
%		& \dfrac{\pi 2^{1-2\lambda}\Gamma(j+2\lambda)}{j!(\lambda + j)[\Gamma(\lambda)]^{2}}, \textrm{if} \ j=k.
%	\end{array}\right.
%\end{eqnarray}

The most general form for the solutions of the Klein-Gordon equation \eqref{eqKG} correspond to the  modes \eqref{sec3freemodessol1}, which allow us to calculate the Wightman function, a necessary element for the computations of the momentum and position dispersions. This quantity will be calculated in the next subsection.

\subsection{Wightman function}\label{SecIIIA}

In order to obtain the positive frequency Wightman function in the Einstein's universe, we first construct the field operator using the general relation \cite{birrell1984quantum}
\begin{eqnarray}\label{sec3fieldoperator}
\hat{\psi}(x) = \sum_{\sigma}[a_{\sigma}\psi_{\sigma}(x) + a_{\sigma}^{\dagger}\psi_{\sigma}^{*}(x)],
\end{eqnarray}
where $\psi_{\sigma}(x)$ are the mode solutions \eqref{sec3freemodessol1} and $x=(t,\chi,\theta,\phi)$. The coefficients $a_{\sigma}^{\dagger}$ and $a_{\sigma}$ are the creation and annihilation operators, respectively, satisfying the standard relation of commutation $[a_{\sigma},a_{\sigma'}^{\dagger}]=\delta_{\sigma\sigma'}$. The summation symbol, in the present case, now holds for the discrete set of quantum numbers $\sigma$ previously defined. Hence, we can obtain the Wightman function using the expression
\begin{eqnarray}\label{sec3wightmanfunction}
\textrm{W}(x,x')&=&\langle 0|\hat{\psi}(x)\hat{\psi}(x') |0\rangle \nonumber \\
&=& \sum_{\sigma}\psi_{\sigma}(x)\psi_{\sigma}^{*}(x').
\end{eqnarray}
In the first line of the above equation we have the average value of the product of two field operators in the vacuum state $|0\rangle$ of the scalar field operator $\hat{\psi}(x)$, defined by Eq. \eqref{sec3fieldoperator}. On the other hand, the second line shows us that we can obtain the Wightman function through the normalized mode solutions \eqref{sec3freemodessol1}, which are scalar functions.

Considering the mode solutions \eqref{sec3freemodessol1}, the correspondent Wightman function is given by Eq. \eqref{sec3wightmanfunction} with the summation symbol defined as
\begin{eqnarray}\nonumber
\sum_{\sigma}\equiv \sum_{n=0}^{\infty}\sum_{\ell=0}^{n}\sum_{m=-\ell}^{\ell},
\end{eqnarray}
so that we arrive at
\begin{eqnarray}\label{sec3eq03}
\textrm{W}(x,x') = \dfrac{1}{4\pi^{2}a_{0}^{3}}\sum_{n=0}^{\infty} \dfrac{(n+1)e^{-i\omega_{n}\Delta t}}{\omega_{n}} \sum_{\ell=0}^{n}&&\dfrac{2^{2\ell}(n-\ell)![\Gamma(\ell+1)]^{2}(2\ell+1)}{\Gamma(n+\ell+2)}\sin^{\ell}(\chi)\sin^{\ell}(\chi')\nonumber\\
&&\times C_{n-\ell}^{\ell+1}(\cos(\chi))C_{n-\ell}^{\ell+1}(\cos(\chi'))P_{\ell}(\cos(\gamma)),
\end{eqnarray}	
where $\Delta t = (t-t')$ and
\begin{eqnarray}\label{prop_add_theo_sphe_harm2}
\cos(\gamma) = \cos(\theta)\cos(\theta') + \sin(\theta)\sin(\theta')\cos(\phi - \phi').
\end{eqnarray}
Note that, in order to obtain the above expression for W$(x,x')$, we use the addition theorem for spherical harmonics \cite{arfken2005mathematical}. The parameter $\gamma$ corresponds to the separation angle between two vectors oriented by the pair of angular coordinates $(\theta, \phi)$ and $(\theta^{\prime}, \phi^{\prime})$, with modules $r$ and $r^{\prime}$, in the spherical coordinate system \cite{jackson1998classical,arfken2005mathematical}.

By using the summation theorem for the Gegenbauer polynomials \cite{gradshtein2007} we can simplify Eq. \eqref{sec3eq03} such that we obtain
\begin{eqnarray}\label{sec3eq04a}
\textrm{W}(x,x') = \dfrac{1}{4\pi^{2}a_{0}^{3}}\sum_{n=0}^{\infty}\dfrac{(n+1)e^{-i\omega_{n}\Delta t}}{\omega_{n}} C_{n}^{1}(\cos(\alpha)),
\end{eqnarray}	
where based on the structure of the angular separation in the relation of the spherical harmonics, that is, in analogy to Eq. \eqref{prop_add_theo_sphe_harm2}, we identify
\begin{eqnarray}\label{sec3eq04b}
\cos(\alpha) = \cos(\chi)\cos(\chi')+\sin(\chi)\sin(\chi')\cos(\gamma).
\end{eqnarray}
The parameter $\alpha$ corresponds to the angular separation between two vectors defined by angular coordinates $(\chi,\theta,\phi)$ and $(\chi^{\prime},\theta^{\prime},\phi^{\prime})$, which can be written in terms of the constant radius $a_{0}$ and the ``spatial'' separation $\Delta s$ according to the relation $\alpha = \Delta s/ a_{0}$. \cite{Ozcan:2006jn,Ozcan:2001cr}. 
%
%In Fig.\ref{figSHaddtheorem}b we illustrate this with the particular case $\phi=\phi^{\prime}$ which allow us to have a pedagogical image of the situation, but we emphasize that this relation holds for arbitrary values of the angular coordinates $\chi$, $\theta$ and $\phi$.
%

In the Einstein's universe, characterized by the line element \eqref{metricEinsteinUniverse}, the Ricci scalar is $R=6a_{0}^{-2}$ and the conformal symmetry provide $\xi =1/6$. Furthermore, from the properties of the Gegenbauer polynomials it is observed that \cite{gradshtein2007}
\begin{eqnarray}\nonumber
C^{1}_{n}(\cos(\alpha)) = \dfrac{\sin[(n+1)\alpha]}{\sin(\alpha)}.
\end{eqnarray}
Then, redefining the summation index, we obtain 
\begin{eqnarray}\label{sec3eq05}
\textrm{W}(x,x') = \dfrac{1}{4\pi^{2}a_{0}^{2}\sin(\alpha)}\sum_{k=1}^{\infty}\dfrac{k\sin(k\alpha)}{\sqrt{k^{2}+a_{0}^{2}m_{F}^{2}}}e^{-i\Delta\tau\sqrt{k^{2}+a_{0}^{2}m_{F}^{2}}},
\end{eqnarray}
where we have defined $\Delta\tau = \Delta t/a_{0}$. The above summation can be computed by using the Abel-Plana formula \cite{Saharian:2007ph}
\begin{eqnarray}\label{prop_abel_plana}
\sum_{k=0}^{\infty} F(k) = \dfrac{1}{2}F(0) + \int_{0}^{\infty}dr F(r) + i\int_{0}^{\infty}dr \dfrac{[F(ir)-F(-ir)]}{\left(e^{2\pi r}-1\right)},
\end{eqnarray}
where in this case we identify 
\begin{eqnarray}\label{sec3eq06}
F(k) \equiv \dfrac{k\sin(k\alpha)}{\sqrt{k^{2}+a_{0}^{2}m_{F}^{2}}}e^{-i\Delta\tau\sqrt{k^{2}+a_{0}^{2}m_{F}^{2}}}.
\end{eqnarray}	
Furthermore, observing that $F(0)=0$ and using the identity
\begin{eqnarray}\nonumber
\sqrt{(\pm ir)^{2}+a_{0}^{2}m_{F}^{2}} =
	\left\{\begin{array}{ll}
				&\sqrt{a_{0}^{2}m_{F}^{2}-r^{2}}, \ \textrm{if}\ r < a_{0}m_{F},\\
				&(\pm i)\sqrt{r^{2}-a_{0}^{2}m_{F}^{2}}, \ \textrm{if}\ r > a_{0}m_{F},
		   \end{array}\right.
\end{eqnarray}
by substituting Eq. \eqref{sec3eq06} into \eqref{prop_abel_plana}, after some algebraic work, we have for Eq. \eqref{sec3eq05} that
\begin{eqnarray}\label{sec3eq07a}
\textrm{W}(x,x') = \textrm{W}_{0}(x,x') + \textrm{W}_{1}(x,x'),
\end{eqnarray}
where for practical purpose we have defined
\begin{eqnarray}\label{sec3eq07b}
\textrm{W}_{0}(x,x') = \dfrac{1}{4\pi^{2}a_{0}^{2}\sin(\alpha)}\int_{0}^{\infty}dr\dfrac{r\sin(r\alpha)}{\sqrt{r^{2}+r_{0}^{2}}}e^{-i\Delta\tau\sqrt{r^{2}+r_{0}^{2}}}
\end{eqnarray}	
and
\begin{eqnarray}\label{sec3eq07c}
\textrm{W}_{1}(x,x') = \dfrac{-1}{2\pi^{2}a_{0}^{2}\sin(\alpha)}\int_{r_{0}}^{\infty}dr\dfrac{r\sinh(r\alpha)}{(e^{2\pi r}-1)}\frac{\cosh(\Delta\tau\sqrt{r^{2}-r_{0}^{2}})}{\sqrt{r^{2}-r_{0}^{2}}},
\end{eqnarray}
with $r_{0}=a_{0}m_{F}$.

All integrations in Eqs. \eqref{sec3eq07b} and \eqref{sec3eq07c} can be calculated with the help of Refs. \cite{gradshtein2007}, \cite{prudnikov1986integralsV1} and \cite{prudnikov1986integralsV2}, such that we obtain
\begin{eqnarray}\label{sec3eq08a}
\textrm{W}_{0}(x,x') = \dfrac{im_{F}}{8\pi a_{0}\sin(\alpha)}\dfrac{\Delta s}{\sqrt{\Delta t^{2} - \Delta s^{2}}}H_{1}^{(2)}(m_{F}\sqrt{\Delta t^{2}-\Delta s^{2}})
\end{eqnarray}
and
\begin{eqnarray}\label{sec3eq08b}
\textrm{W}_{1}(x,x') = \dfrac{im_{F}}{8\pi a_{0}\sin(\alpha)}\sideset{}{'}\sum_{n=-\infty}^{\infty}\dfrac{(\Delta s +2\pi a_{0} n)}{\sqrt{\Delta t^{2}-(\Delta s +2\pi a_{0} n)^{2}}}H_{1}^{(2)}(m_{F}\sqrt{\Delta t^{2}-(\Delta s +2\pi a_{0} n)^{2}}).
\end{eqnarray}	
In Eq. \eqref{sec3eq08b}, the prime symbol indicates that the $n=0$ term is not included in the summation. In order to write W$_{0}(x,x')$ and W$_{1}(x,x')$ in terms of the Hankel function or Bessel function of the third kind $H_{1}^{(2)}(z)$ we have used the relation $K_{1}(iz)=(-\pi/2)H_{1}^{(2)}(z)$, where $K_{\nu}(z)$ is known as Macdonald function \cite{prudnikov1986integralsV2, gradshtein2007}. Moreover, in order to obtain W$_{1}(x,x')$ we have used the exponential representation for the hyperbolic sine function and also \cite{Saharian:2007ph}
$$
\dfrac{1}{(e^{2\pi r}-1)} = \sum_{n=1}^{\infty}e^{-(2\pi r)n}.
$$

Finaly from the Eqs. \eqref{sec3eq07a}, \eqref{sec3eq08a} and \eqref{sec3eq08b} we can write
\begin{eqnarray}\label{eqWF_EinsteinUniverseA}
\textrm{W}(x,x') = \dfrac{im_{F}}{8\pi a_{0}\sin\left(\frac{\Delta s}{a_{0}}\right)}\sum_{n=-\infty}^{\infty}\dfrac{(\Delta s +2\pi a_{0} n)}{\sigma_{n}}H_{1}^{(2)}(m_{F}\sigma_{n}),
\end{eqnarray}
where as we know $m_{F}$ is the field mass, $a_{0}$ the Einstein universe constant radius and $\sigma_{n}$ the spacetime separation vector defined as
\begin{eqnarray}\label{eqWF_EinsteinUniverseB}
\sigma_{n}^{2} = \Delta t^{2} - (\Delta s+2\pi a_{0} n)^{2}.
\end{eqnarray}
Note that the $n=0$ term corresponds to the analogue of the Minkowski vacuum contribution, which come from the W$_{0}(x,x')$ integral, Eq. \eqref{sec3eq08a}. It is important to stress that although the structure of the contribution $n=0$ in the Einstein's universe is not equal to the unbounded Minkowski vacuum contribution, in the limit $a_{0}\rightarrow\infty$ the Einstein's universe with finite size indeed becomes the infinite-sized Minkowski spacetime contribution \cite{dowker1977vacuum,dowker1976covariant}.

Eq. \eqref{eqWF_EinsteinUniverseA} corresponds to the expression for the positive frequency Wightman function of a massive scalar field in the Einstein's universe. Although it provides a more realistic description of the model, that is, with more details about influences of each of the elements involved, its general structure increases the difficulty in mathematical calculations. Therefore, in a preliminary analysis, and for the sake of simplicity, it is instructive to first consider the massless scalar field case. Taking the limit $m_{F}\rightarrow 0$ in Eq. \eqref{eqWF_EinsteinUniverseA} we have \cite{dowker1976covariant}
\begin{eqnarray}\label{eqWF_EinsteinUniverseC}
\textrm{W}(x,x') = - \dfrac{1}{4a_{0}\pi^{2}}\sum_{n=-\infty}^{\infty}\dfrac{(\Delta s +2\pi a_{0}n)}{\sin\left(\frac{\Delta s}{a_{0}}\right)\sigma_{n}^{2}},
\end{eqnarray}
where all parameters have already been defined previously. Eq. \eqref{eqWF_EinsteinUniverseC} corresponds to the positive frequency Wightman function for a massless scalar field in the Einstein's universe.
It is important to mention that there is a different version of Eq. \eqref{eqWF_EinsteinUniverseC}, in which the summation is not present. In fact, taking the massless limit in Eq. \eqref{sec3eq05} it can be shown that \cite{Ozcan:2001cr, Ozcan:2006jn}
\begin{eqnarray}\label{eqWF_EinsteinUniverseD}
\textrm{W}(x,x')=\dfrac{1}{8a_{0}^{2}\pi^{2}}\dfrac{1}{\left[\cos\left(\frac{\Delta t}{a_{0}}\right)-\cos\left(\frac{\Delta s}{a_{0}}\right)\right]}.
\end{eqnarray}
Different from Eq. \eqref{eqWF_EinsteinUniverseC}, no summation is present in Eq. \eqref{eqWF_EinsteinUniverseD}. Although both expressions are equivalent, for our purposes, Eq. \eqref{eqWF_EinsteinUniverseC} is more convenient since it allows us to extract directly the divergent term ($n=0$), in order to regularize our results. In contrast, the structure of Eq. \eqref{eqWF_EinsteinUniverseD} does not allow us to easily see how to perform such a procedure in order to eliminate the divergent contribution.

In the next sections we will use Eq. \eqref{eqWF_EinsteinUniverseD} to obtain and study the behavior of the momentum and position dispersions induced on a point particle by the quantum vacuum fluctuation of a massless scalar field in the Einstein's universe.

\section{Momentum and position dispersions}\label{Sec3}

Now we will establish the necessary expressions to calculate the dispersion in the momentum and position of a point particle, caused by its interaction with a quantum fluctuating massless scalar field that pervades the spacetime defined by Einstein's universe \eqref{metricEinsteinUniverse}. Initially, we introduce the dynamics of a point particle in curved space time and obtain the classical expressions from which, through the quantization prescription method ($\psi\rightarrow\hat{\psi}$, $p\rightarrow\hat{p}$ and $x\rightarrow\hat{x}$), we obtain the expressions for the dispersion in the momentum and position of the particle. 

\subsection{General expressions and particle dynamics}\label{Sec_Gen_exp}

The dynamics of a point particle of mass $m_{\textrm{p}}$ and charge $q$ coupled to a massless scalar field $\psi(z)$ in a curved spacetime is determined by \cite{poisson2011motion,bessa2017quantum,mota2020induced}
\begin{eqnarray}\label{sec4eq00}
m_{\textrm{p}}(\tau)\dfrac{Du^{\mu}}{d\tau} = q\left(-g^{\mu\nu}+u^{\mu}u^{\nu}\right)\nabla_{\nu}\psi(x),
\end{eqnarray}
where $u^{\mu}=dx^{\mu}/d\tau$ is the four-velocity of the particle, $\tau$ is the proper time and $x^{\mu}$ stands for the set of spacetime coordinates. The mathematical object $Du^{\mu}/d\tau$ corresponds to the covariant derivative for the components of the four-velocity vector $u^{\mu}$. 
%
%, given by 
%
%\begin{eqnarray}\label{sec4eq00a}
%\dfrac{Du^{\mu}}{d\tau} = \dfrac{du^{\mu}}{d\tau} +\Gamma^{\mu}_{\nu\rho}u^{\nu}u^{\rho},
%\end{eqnarray}
%
%where the coefficients $\Gamma^{\mu}_{\nu\rho}$ are the Christoffel symbols, which can be obtained by means of relation \eqref{sec4eq00b}. 
%
Note that since $\psi(x)$ is a scalar field, in Eq. \eqref{sec4eq00}, $\nabla_{\nu}\psi(x)=\partial_{\nu}\psi(x)$ \cite{hobson2006}.

Once a point particle has its dynamics description in curved spacetime it will radiate energy, producing variation in its rest mass. In other words, the point particle mass is time dependent \cite{poisson2011motion}. The variation of the dynamical mass $m_{\textrm{p}}(\tau)$ is described by the first order differential equation 
\begin{eqnarray}\label{sec4massvariationa}
\dfrac{dm_{\textrm{p}}(\tau)}{d\tau} = -qu^{\mu}\nabla_{\mu}\psi(x),
\end{eqnarray}
which admits the linear solution
\begin{eqnarray}\label{sec4massvariationb}
m_{\textrm{p}}(\tau) := m_{0}-q\psi(x),
\end{eqnarray}
where $m_{0}$ is the constant mass of the particle.

In the present study we consider a regime in which the particle's motion is slow enough so that we can assume that spatial coordinates are approximately time independent \cite{yu2004vacuum, bessa2009brownian,seriu2009smearing,seriu2008switching,de2014quantum,de2016probing,camargo2018vacuum}. Thus, in this particular case, proper and coordinate times are equal and from Eqs. \eqref{sec4eq00} and \eqref{sec4massvariationa} we obtain 
\begin{eqnarray}\label{sec4eq01}
\dfrac{dp^{i}}{dt} +m_{\textrm{p}}\Gamma^{i}_{\alpha\beta}u^{\alpha}u^{\beta} = -qg^{ij}\nabla_{j}\psi(x) + f^{i}_{\textrm{ext}},
\end{eqnarray}
where $p^{i} = m_{\textrm{p}}(t)u^{i}(t)$ is the spatial component of the particle's momentum. It is important to note that all quantities in the expression above are still classical since we have not so far implemented any process of quantization of the physical observables. Furthermore, we have considered the extra term $f_{\textrm{ext}}^{i}$ in order to include possible external and classical contributions to the point particle dynamics.

%{\color{blue} Aqui, os efeitos de backreaction também serão negligenciados tendo em vista o regime não relativístico da abordagem e o caráter de partícula pontual (or test particle) assumido.}

Here, the backreaction effects are neglected. In curved spacetimes, that is, in the presence of gravity, backreaction effects are a natural consequece. A curved spacetime modifies the quantum vacuum fluctuations of the fields and consequently this provides a nonzero renormalized vacuum expectation value (VEV) for the energy-momentum tensor \cite{mukhanov2007introduction}. Then, according to Einstein's field equations, this nonzero energy-momentum tensor is also a field source and thus modifies the classical geometry of the spacetime. This effect of the geometric modification, resulting from vacuum fluctuations, can be encoded in the metric tensor associated with the spacetime. In this direction, for instance, in Ref. \cite{DeLorenci:2008nr} the authors obtained up to order $\hbar$ the quantum correction for the metric tensor of a spinning cosmic string, due to the backreaction effects from the renormalized VEV of the energy-momentum tensor of a conformally coupled massless scalar field. In a distinct context, the influences of backreaction effects in Einstein's universe were discussed in Refs. \cite{Altaie:2001vv} and \cite{Altaie:2002tv}, considering a conformally coupled massless scalar field, a photon field and a neutrino field, at finite temperature.

Using Eq. \eqref{metricEinsteinUniverse} we find that the only non vanishing Christoffel symbols are those shown in Table \ref{tableCSEuniverse}. From these results we see that solving \eqref{sec4eq01} is a hard task due to the coupling of the distinct components of velocity and momentum in the general expression. However, the contributions from the terms proportional to the coefficients $\Gamma^{i}_{\alpha\beta}$ can be interpreted as classical fictitious forces \cite{adler2021general}. So, as these coefficients are of geometric origin it is plausible to identify 
\begin{eqnarray}\label{sec4eq02}
f_{\textrm{ext}}^{i}=m_{\textrm{p}}\Gamma^{i}_{\alpha\beta}u^{\alpha}u^{\beta}.
\end{eqnarray}
In this approach we are regarding that quantum contributions come exclusively from the massless scalar field and are not related to geometric aspects of space. In other words, the geometry is classical and can only modify the quantum effects coming from the scalar field. 

In order to maintain clarity, let us now further discuss the choice for the external force in Eq. \eqref{sec4eq02}. After we quantize the physical observables in Eq. \eqref{sec4eq01} we note that the Einstein's universe geometry considered in this work affects the IQBM of the particle in two distinct ways. Through the escalar field modes, which interact with the point particle, and by means of the coefficients $\Gamma_{\alpha\beta}^{i}$ and $g^{ij}$ present in Eq. \eqref{sec4eq01}. Thus, in principle, a more complete approach is reached by taking into account the Christoffel symbols as well, in which case the external force would be set as being zero. However, this scenario leads to technical problems that do not allow us to analytically solve Eq. \eqref{sec4eq01} without an additional assumption, as the one in Eq. \eqref{sec4eq02}. Therefore, the choice of the latter makes possible to analytically solve Eq. \eqref{sec4eq01} and perform a full analysis of the IQBM, as we shall see below.

We can understand the meaning of the external force introduced in Eq. \eqref{sec4eq02} as a way of making the particle to feel the effects of the background geometry only through the induced quantum vacuum fluctuations of the scalar field.  In other words, our approach assumes that the geometry of the spacetime (a local property) has little influence on the particle's equation of motion and, thus, on the particle's dynamics. However, the field, which occupies all of the spacetime, probes the entire geometric structure under consideration which in turn affects the propagation of its modes. Thus, although we are neglecting a potential contribution to the IQBM, with the choice of Eq. \eqref{sec4eq02}, we can still study the influences of the spacetime through the quantum vacuum fluctuations of the scalar field, which probes the nontrivial geometric structure of space and transmits this information to the particle. We can find something similar in the literature, for example, in Refs. \cite{bessa2009brownian} e \cite{bessa2017quantum}, where a non-fluctuating (classical) external force cancels the effects of the spacetime expansion.
\begin{table}[h]
\caption{Non-zero Christoffel symbols for Einstein's universe.}\label{tableCSEuniverse}
	\begin{tabular}{cc}
	\toprule
	 	$\Gamma^{\chi}_{\theta\theta}$ & $-\sin(\chi)\cos(\chi)$ \\ \midrule
	  	$\Gamma^{\chi}_{\phi\phi}$ & $-\cos(\chi)\sin(\chi)\sin^{2}(\theta)$ \\ \midrule
	  	$\Gamma^{\theta}_{\chi\theta}, \Gamma^{\theta}_{\theta\chi},\Gamma^{\phi}_{\chi\phi},\Gamma^{\phi}_{\phi\chi}$ & $\cot(\chi)$ \\ \midrule
	  	$\Gamma^{\theta}_{\phi\phi}$ & $-\sin(\theta)\cos(\theta)$ \\ \midrule
	  	$\Gamma^{\phi}_{\theta\phi},	\Gamma^{\phi}_{\phi\theta}$ & $\cot(\theta)$\\
	\bottomrule
	\end{tabular}
\end{table}

Now, considering Eqs. \eqref{sec4eq01} and \eqref{sec4eq02} we can obtain the following expression for the particle's momentum:
\begin{eqnarray}\label{sec4eq03}
p^{i}(x) = - q\int_{0}^{\tau}dtg^{ij}\nabla_{j}\psi(x),
\end{eqnarray}
where we have assumed a null initial momentum value, $p^{i}(t=0)=0$. In this expression $\tau$ is an arbitrary constant value of time. In addition, we observe that since $p^{i}(x) = m_{\textrm{p}}(\tau)u^{i}(x)$ we can easily obtain an expression for the velocity of the particle.

In order to obtain the momentum dispersion induced by the quantum fluctuations of $\hat{\psi}$ in the vacuum state $|0\rangle$ we must first quantize Eq. \eqref{sec4eq03}. For this we use a prescription process in which we promote the classical scalar field to a field operator, in other words, the classical field $\psi$ is replaced by a quantum field operator $\hat{\psi}$, which follows the construction shown in Eq. \eqref{sec3fieldoperator}. Then, implementing the described quantization process, the general expression for the dispersion in the momentum components will be given by  
\begin{eqnarray}\label{sec4eq05}
\langle(\Delta \hat{p}^{i})^{2}\rangle_{\textrm{ren}} &=& \langle(\hat{p}^{i})^{2}\rangle - \langle\hat{p}^{i}\rangle^{2}\nonumber\\
&=&\lim_{x'\rightarrow x}\left[ \langle \hat{p}^{i}(x)\hat{p}^{i}(x')\rangle -  \langle \hat{p}^{i}(x)\hat{p}^{i}(x')\rangle_{\textrm{div}}\right],
\end{eqnarray}
where $\langle\ldots\rangle\equiv\langle 0|\ldots|0\rangle$. In the above equation we have used the fact that $\langle \hat{p}^{i}(x)\rangle = 0$, a result which is consequence of the linear relation between the particle momentum and field operator, as shown in \eqref{sec4eq03}, since $a|0\rangle = 0$ and $\langle 0|a^{\dagger} = 0$. Hence, in this case, we notice that the dispersion and the mean value in the vacuum state for the squared particle momentum are equivalents, that is, $\langle(\Delta \hat{p}^{i})^{2}\rangle=\langle(\hat{p}^{i})^{2}\rangle$.

To obtain the result \eqref{sec4eq05} it is important to note that we also use a regularization procedure in order to renormalize (ren) the observable $\langle(\Delta \hat{p}^{i})^{2}\rangle$. For this purpose, we subtract the term $n = 0$ from the Wightman function \eqref{eqWF_EinsteinUniverseC}, which is the only divergent (div) term in the coincidence limit $(\Delta t, \Delta s)\rightarrow(0, 0)$ \cite{Ozcan:2006jn, dowker1977vacuum,dowker1976covariant}. In fact, divergences are typical of Quantum Field Theory and, as it is known, a regularization procedure must be used in order to identify and remove by means of renormalization existing divergences, making possible to find a finite result in the coincidence limit \cite{birrell1984quantum}. Although there are several procedures through which one can perform the process of regularization and renormalization of infinities, the most convenient one chosen here is the point-splitting method \cite{Ozcan:2006jn}. In the present study we consider a curved spacetime, but similar to Refs. \cite{Ozcan:2006jn, dowker1977vacuum,dowker1976covariant} the renormalization procedure used here consist simply in subtracting the contribution $n = 0$.

From Eqs. \eqref{sec4eq03} and \eqref{sec4eq05}, the renormalized momentum dispersion for the point particle will be given by the general expression
\begin{eqnarray}\label{sec4momentum_disp_gen_relation}
\langle(\Delta \hat{p}^{i})^{2}\rangle_{\textrm{ren}} = \lim_{x'\rightarrow x}\dfrac{q^{2}}{2}\int_{0}^{\tau}dt'\int_{0}^{\tau}dt g^{ii}(x)g^{ii}(x')\dfrac{\partial^{2}G_\textrm{ren}^{(1)}(x,x')}{\partial x^{i}\partial x'^{i}},
\end{eqnarray}
where $i=(\chi,\theta,\phi)$ specifies the momentum components and $g^{ii}(x)$ the contravariant components of the metric tensor. Note that, we have also used the fact that the metric tensor is diagonal. The renormalized Hadamard's function $G_\textrm{ren}^{(1)}(x,x')$ present in the above expression arises from the symmetrization of the fields product and can be obtained from the positive frequency Wightman function by means of the relation $G^{(1)}(x,x')=2\textrm{I\!ReW}(x,x')$ \cite{fulling1989aspects}. It is worth mentioning that, as indicated in Eq. \eqref{sec4momentum_disp_gen_relation}, we have already subtracted the divergent contribution coming from $n=0$ which means that we can take the coincidence limit $x=x'$ whenever it is convenient. From now on, we will drop the use of the limit, leaving it implied.

Before ending the present subsection we would like to briefly point out an interesting result: the dynamical mass can fluctuate. In our semiclassical approach, the structure of the expression for the dynamical mass, Eq. \eqref{sec4massvariationb}, shows that in the quantization process the mass becomes an operator. Its average value in the vacuum state exactly corresponds to the constant mass, $\langle \hat{m}_{\textrm{p}}\rangle = m_{0}$. In addition, we can also obtain the mean value of the renormalized squared mass $\langle \hat{m}_{\textrm{p}}^{2}\rangle_{\textrm{ren}}$ and, consequently, the mass dispersion $\langle(\Delta\hat{m}_{\textrm{p}})^{2}\rangle_{\textrm{ren}}$. 
In fact, from Eqs. \eqref{eqWF_EinsteinUniverseC} and \eqref{sec4massvariationb} we can show that in the coincidence limit 
$$\langle(\hat{m}_{\textrm{p}})^{2}\rangle_{\textrm{ren}}=m_{0}^{2}+q^{2}\langle\hat{\psi}^{2}\rangle_{\textrm{ren}}$$ 
and, consequently, 
$$\langle(\Delta\hat{m}_{\textrm{p}})^{2}\rangle_{\textrm{ren}}=q^{2}\langle\hat{\psi}^{2}\rangle_{\textrm{ren}},$$ 
where 
$$\langle\hat{\psi}^{2}\rangle_{\textrm{ren}}=\lim_{x'\rightarrow x}\textrm{W}_{\textrm{ren}}(x,x')=-\dfrac{1}{48\pi^{2}a_{0}^{2}}$$
is the mean value for the squared field in the vacuum state. In the limit $a_{0}\rightarrow\infty$, restoring Minkowski spacetime, we notice that $\langle(\hat{m}_{\textrm{p}})^{2}\rangle_{\textrm{ren}}=m_{0}^{2}$ and $\langle(\Delta\hat{m}_{\textrm{p}})^{2}\rangle_{\textrm{ren}}=0$, indicating that the mass does not fluctuate. Also, we note that $\langle(\Delta\hat{m}_{\textrm{p}})^{2}\rangle_{\textrm{ren}}<0$ and this peculiar result, at first glance, seems strange, since the dispersion is a positive quantity. However, this is another issue in calculating the mean value of observables (in the vacuum state) in Quantum Field Theory, where it is also possible to obtain negative results for the mean value of quadratic quantities. In the literature, this fact is known as being due to subvacuum effects. See for example Refs. \cite{DeLorenci:2018moq} and \cite{Wu:2008am}. As pointed out in Ref. \cite{fulling1989aspects} this can be understood, for instance, as a consequence of the renormalization process.

In the next subsection, we will use Eq. \eqref{sec4momentum_disp_gen_relation} and the results of Section \ref{SecIIIA} to calculate and analyze the behavior of the dispersion in the momentum components. 

\subsection{Momentum component dispersion}\label{Sec4_momentum_dispertions}

Using all the results and formalism shown in the preceding sections, we can now calculate the dispersion for the components of the particle's momentum in the Einstein's universe. According to Eq. \eqref{sec4momentum_disp_gen_relation} the algorithm consists of choosing a componente $i$ and identifying the corresponding elements of the contravariant metric tensor $g^{ii}(x)$ from Eq. \eqref{metricEinsteinUniverse}. Next, we perform the derivatives and integrals operations and analyze the results.
	
Following the steps described above, for the angular component $i=\chi$ we obtain that
\begin{eqnarray}\label{sec4eq06}
\langle(\Delta \hat{p}^{\chi})^{2}\rangle_{\textrm{ren}} = 2q^{2}a_{0}^{-4}\int_{0}^{\tau}d\eta(\tau-\eta) K_{\chi}(x,x'),
\end{eqnarray}
where we have used the identity \cite{de2014quantum, camargo2018vacuum}
\begin{eqnarray}\label{sec4eq_integral_identity}
\int_{0}^{\tau}dt'\int_{0}^{\tau}dt\mathcal{G}(|t-t'|) = 2\int_{0}^{\tau}d\eta(\tau-\eta)\mathcal{G}(\eta), 
\end{eqnarray}
with $\eta=|t-t'|$ and also defined 
\begin{eqnarray}\label{sec4eq_kernel}
K_{i}(x,x')=\partial_{i}\partial_{i'}\textrm{W}_{\textrm{ren}}(x,x').
\end{eqnarray}	
As it is clear from Eqs. \eqref{sec4eq06} and \eqref{sec4eq_kernel}, for each component $i$ we have the integral
\begin{eqnarray}\label{sec4eq_integral_kernel}
I_{i}(x,x')=\int_{0}^{\tau}d\eta(\tau-\eta)K_{i}(x,x').
\end{eqnarray}	

For the $\chi$ component of the particle's momentum, using Eqs. \eqref{eqWF_EinsteinUniverseC}, \eqref{sec4eq06}, \eqref{sec4eq_kernel} and \eqref{sec4eq_integral_kernel}, we find that the dispersion in the coincidence limit will be given by	
\begin{eqnarray}\label{sec4eq_mdisp_chi}
\langle(\Delta \hat{p}^{\chi})^{2}\rangle_{\textrm{ren}} = 2q^{2}a_{0}^{-4}I_{\chi}(\tau_{a}),
\end{eqnarray}
where we have defined the quantity
\begin{eqnarray}\label{sec4eq_mdisp_chi_F}
I_{\chi}(\tau_{a}) = - \dfrac{1}{(12\pi)^{2}}\left\{1+\frac{12}{\tau_{a}^{2}}-3\csc^{2}\left(\dfrac{\tau_{a}}{2}\right) +6\ln\left[\dfrac{\sin\left(\frac{\tau_{a}}{2}\right)}{\left(\frac{\tau_{a}}{2}\right)}\right]^{2}  \right\},
\end{eqnarray}	
and the dimensionless time parameter $\tau_{a}=\tau/a_{0}$. In order to clarify the attainment of the above result, before proceeding, let us outline the methodology used. To calculate the contribution $I_{\chi}$, we first have performed the sum and taken in advance the coincidence limit in the variables $\theta$ and $\phi$, that is, $(\theta',\phi')\rightarrow(\theta,\phi)$, since the operations can only affect the coordinates $\chi$ and $\chi'$. Then, we have derived the resulting expression with respect to the variables $\chi $ and $\chi'$, in addition to taking the limit $\chi=\chi'$ at the end. Next, we compute the integral \eqref{sec4eq_integral_kernel} using $K_{\chi}$ to find the results shown in Eqs. \eqref{sec4eq_mdisp_chi} and \eqref{sec4eq_mdisp_chi_F}.

For the theta component of momentum dispersion, taking $i=\theta$ in Eq. \eqref{sec4momentum_disp_gen_relation}, we obtain
\begin{eqnarray}\label{sec4eq07}
\langle(\Delta \hat{p}^{\theta})^{2}\rangle_{\textrm{ren}} = 2q^{2}a_{0}^{-4}\sin^{-4}(\chi)\int_{0}^{\tau}d\eta(\tau-\eta) K_{\theta}(x,x'),
\end{eqnarray}
where we have used the identity \eqref{sec4eq_integral_identity} and the definition \eqref{sec4eq_kernel}. By computing the integral for $K_{\theta}$ as defined in Eq. \eqref{sec4eq_integral_kernel}, we find that
\begin{eqnarray}\label{sec4eq_mdisp_theta}
\langle(\Delta \hat{p}^{\theta})^{2}\rangle_{\textrm{ren}} = 2q^{2}a_{0}^{-4}\sin^{-4}(\chi) I_{\theta}(\chi,\tau_{a}),
\end{eqnarray}
with
\begin{eqnarray}\label{sec4eq_mdisp_theta_F}
I_{\theta}(\chi,\tau_{a})=\sin^{2}(\chi)I_{\chi}(\tau_{a}).
\end{eqnarray}
To solve the integrals $I_{\theta}$ we have followed a similar procedure to that described for the component $\chi$. From Eq. \eqref{sec4eq_mdisp_theta_F} we also note that the theta component is related to the contribution of the $\chi$ component, Eq. \eqref{sec4eq_mdisp_chi_F}, and is modulated by an amplitude that depends on the angular variable $\chi$.

Finally, for the $i=\phi$ component, from Eq. \eqref{sec4momentum_disp_gen_relation}, we have
\begin{eqnarray}\label{sec4eq08}
\langle(\Delta \hat{p}^{\phi})^{2}\rangle_{\textrm{ren}} = 2q^{2}a_{0}^{-4}\sin^{-4}(\chi)\sin^{-4}(\theta)\int_{0}^{\tau}d\eta(\tau-\eta) K_{\phi}(x,x').
\end{eqnarray}
Using all the mathematical techniques and manipulations applied in the previous component calculations, we can calculate the above integral and show that
\begin{eqnarray}\label{sec4eq_mdisp_phi}
\langle(\Delta \hat{p}^{\phi})^{2}\rangle_{\textrm{ren}} = 2q^{2}a_{0}^{-4}\sin^{-4}(\chi)\sin^{-4}(\theta)I_{\phi}(\theta,\chi,\tau_{a}),
\end{eqnarray}
with
\begin{eqnarray}
I_{\phi}(\theta,\chi,\tau_{a})=\sin^{2}(\theta)I_{\theta}(\chi,\tau_{a})=\sin^{2}(\theta)\sin^{2}(\chi)I_{\chi}(\tau_{a})
\end{eqnarray}
where $I_{\theta}(\chi,\tau_{a})$ and $I_{\chi}(\tau_{a})$ are defined in Eqs. \eqref{sec4eq_mdisp_theta_F} and \eqref{sec4eq_mdisp_chi_F}, respectively. 

\begin{figure}[h]
\includegraphics[scale=0.5]{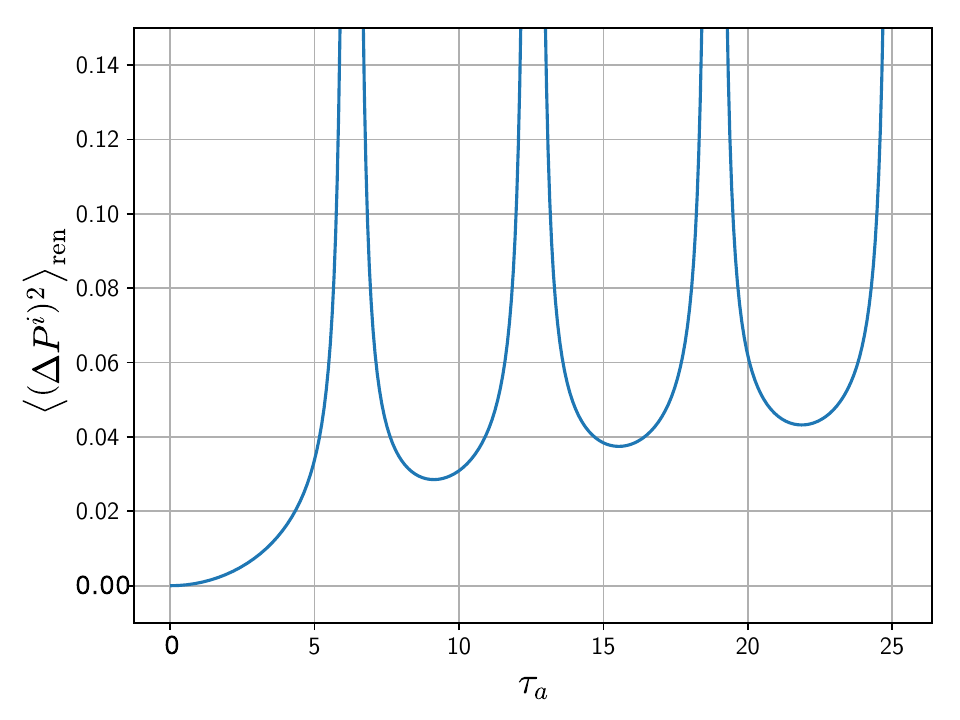}
\caption{Behavior of the renormalized dimensionless dispersion of physical momentum for a point particle coupled to a real massless scalar field in the Einstein's universe, as a function of the dimensionless time $\tau_{a}$, for the components $\chi$, $\theta$ and $\phi$. Here, for simplicity, we have defined $\langle(\Delta P^{i})^{2}\rangle_{\textrm{ren}} = \left(\frac{a_{0}}{q}\right)^{2}\langle(\Delta \hat{\mathcal{p}}^{i})^{2}\rangle_{\textrm{ren}}$, with $i=\chi,\theta,\phi$.}\label{figMomentumDispersion}
\end{figure}

Eqs. \eqref{sec4eq_mdisp_chi}, \eqref{sec4eq_mdisp_theta} and \eqref{sec4eq_mdisp_phi} correspond to the expressions for the renormalized dispersion of the momentum components. To obtain the dispersions referrings to the physical momentum, $\mathcal{p}^{i}$, we use the relations
\begin{eqnarray}\nonumber
\mathcal{p}^{i}=\{\mathcal{p}^{\chi} ; \mathcal{p}^{\theta} ; \mathcal{p}^{\phi}\}=\{a_{0}p^{\chi} ; a_{0}\sin(\chi)p^{\theta} ; a_{0}\sin(\chi)\sin(\theta)p^{\phi}\},
\end{eqnarray}
%
%$\mathcal{p}^{\chi}=a_{0}p^{\chi}$, $\mathcal{p}^{\theta}=a_{0}\sin(\chi)p^{\theta}$ and $\mathcal{p}^{\chi}=a_{0}\sin(\chi)\sin(\theta)p^{\phi}$, 
%
which can be deduced from the metric in Eq. \eqref{metricEinsteinUniverse}. Therefore, using the appropriate relations shown above, we find that the dispersions of the particle's renormalized physical momentum will be given by general relation
\begin{eqnarray}\label{sec4eq09}
\langle(\Delta \hat{\mathcal{p}}^{i})^{2}\rangle_{\textrm{ren}} = \dfrac{2q^{2}}{a_{0}^{2}}I_{\chi}(\tau_{a}),
\end{eqnarray}
with $i=(\chi,\theta,\phi)$ and $I_{\chi}(\tau_{a})$ given by \eqref{sec4eq_mdisp_chi_F}. This result shows that the mean value for the dispersion of the physical momentum of the particle is the same for all components, in other words, it is homogeneous and isotropic. As can be easily seen from Eq. \eqref{sec4eq09}, except for the constants, the behavior of $\langle(\Delta \hat{\mathcal{p}}^{i})^{2}\rangle_{\textrm{ren}}$ is similar to that of the function $I_{\chi}(\tau_{a})$ and is duly shown in Fig.\ref{figMomentumDispersion}.

The homogeneous and isotropic results shown in \eqref{sec4eq09} are understandable, since FRW universe is homogeneous and isotropic in large scale. Therefore, Einstein's universe, which corresponds to the particular case $k=+1$, with constant scale factor, also exhibits such properties through the observable $\langle(\Delta \hat{\mathcal{p}}^{i})^{2}\rangle_{\textrm{ren}}$. In Ref. \cite{bessa2017quantum} a similar result was found, in which the authors also obtain an equally homogeneous and isotropic velocity dispersion, considering an analogue model scenario with a Bose-Einstein condensate to simulate a conformal and asymptotically flat expanding universe.

For the limit $\tau_{a}\rightarrow 0$ we note that $\langle(\Delta \hat{\mathcal{p}}^{i})^{2}\rangle_{\textrm{ren}}=0$ and, possibly, this result is a consequence of the classical conditions initially assumed in Eq. \eqref{sec4eq03}, such that $p^{i}(t)=0$ for $t=0$. In the limit $a_{0}\rightarrow\infty$ we also obtain that $\langle(\Delta \hat{\mathcal{p}}^{i})^{2}\rangle_{\textrm{ren}}=0$, which suggests that in Minkowski spacetime there is no IQBM. This is an acceptable results since we work with renormalized observables, that is, quantities whose divergent contributions from Minkowski spacetime, which arise in the coincidence limit, have been subtracted.

Observing the behavior of the mean value of the physical momentum squared  in Fig.\ref{figMomentumDispersion} and its corresponding expression \eqref{sec4eq09}, written in terms of Eq. \eqref{sec4eq_mdisp_chi_F}, we note that there are regular divergences related to the dimensionless time $\tau_{a}$, specifically for dimensional time values  $\tau=(2\pi a_{0})n$, with integer $n\geq 1$. These divergences occur due to the nontrivial topology of Einstein's universe, whose spatial section is closed or compact (S$^{3}$) for all cosmic time value $t$ (represented by I\!R). The global spacetime topology of this universe model, I\!R$^{1}\times$S$^{3}$, is called cylindrical, because in a geometrical representation each cross section of the cylinder correspond to a compact hypersurface S$^{3}$ defined by a constant cosmic time value \cite{d1992introducing}. Consequently, this works as an effect analogous to that which comes from compactified systems \cite{Ferreira:2023uxs}. 
%
%\sout{In fact, substitution of $r=\sin(\chi)$ in Eq. \eqref{metricFLRW2}, which gives the line element \eqref{metricEinsteinUniverse}, shows the `compactification' effect. That is, similar to a spherical geometry, the radius $r=[0,\infty]$ now can bee seen as compactified $r=\sin(\chi)=[0,1]$, since $\chi=[0,\pi]$ in a hyperspheric geometry.}
%
The divergencies that appear in the present work are also similar to the round trip divergencies arising in systems that consider the effect of two parallel planes \cite{yu2004brownian,yu2004vacuum,de2014quantum}. However, it is important to emphasize that, in the present case, the divergences are related to the compact topology of the spacetime. In fact, here, no material boundary (such as planes or plates) is present to influence the modes of the field. In this sense, the observed divergences are possibly associated with the classical treatment adopted for the geometry of spacetime. A more elaborated approach, treating the geometry quantum mechanically (fluctuating), could perhaps eliminate these divergences. Something similar in this direction occurs in the study of quantum fluctuations of the lightcone, where some classical singularities are shown to be removed on the ligthcone by considering quantum field theory in the linearized framework of general relativity. For more details on this aspect, see Ref. \cite{Ford:1994cr}. Also, it can be expected that backreaction effects (here neglected) may have some influence on the smoothness of these divergences.

In the spacetime geometrically defined by line the element \eqref{metricEinsteinUniverse} the time $\tau=(2\pi a_{0})n$ corresponds to multiple length of circumference defined by $\chi=\theta=\pi/2$, for a fixed time $t$. Therefore, in the present case, the analogous round trip divergences are related to the time in which a light signal performs a complete turn around in a circle with length $\tau=2\pi a_{0}$. In addition, observing Fig.\ref{figMomentumDispersion} we notice that the shape of the curves for $n>1$ are equal, but at each turn around the circumference of length $2\pi a_{0}$ the dispersion becomes increasingly positive. This is a nontrivial behavior and suggest that the point particle has its momentum dispersion increased through a nontrivial physical process. In view of this behavior, we can say that, in principle, if there are subvacuum effects, they are possibly suppressed at each turn.

\subsection{Position dispersion and small displacemente condition}

In order to obtain the results presented in the previous subsection we have considered the hypothesis of the small displacement condition, in other words, that the coordinates variations with respect to time are so small that we can neglect them. This assumption is a simplification and imposes some constraints on the results for the previously analyzed momentum dispersions.  We emhasize that this is a fundamental simplification for the approach we use, since this way it is possible to directly identify the mean value of the field product in the vacuum state as the Wightman function.

Since the changes in the particle's coordinates are small, its average value is very close to the real position and, consequently, the dispersion is very small. Thus, we must obtain the expression for the coordinate dispersion and analyze the necessary constraints that we need to impose on the free parameters in order to satisfy the small displacement condition or, equivalently, maintain the coordinate dispersion very small. 

An expression for the coordinates of the particle can be obtained from Eq. \eqref{sec4eq03} observing that $u^{i}=dx^{i}/dt$:
\begin{eqnarray}
m(t)\dfrac{dx^{i}(t)}{dt} = - q\int_{0}^{\tau}dt g^{ij}\partial_{j}\psi(x).
\end{eqnarray}
Thus, the small displacement condition also allows us to simplify the above expression and write it as
\begin{eqnarray}\label{sec4_eq_position}
x^{i}(\tau)=-\dfrac{q}{m_{0}}\int_{0}^{\tau}dt\int_{0}^{t}dt'g^{ij}\partial_{j}^{'}\psi(x'),
\end{eqnarray}
where we should remember that $m_{0}$ is the constant mass of the particle and we also assume that $x^{i}(t=0)=0$, which is a classical assumption.

Similar to Section \ref{Sec_Gen_exp}, to study quantum fluctuations in the particle's coordinates, we now need to quantize Eq. \eqref{sec4_eq_position} by means of the quantization prescription $\psi\rightarrow\hat{\psi}$, which naturally implies that $x^{i}\rightarrow\hat{x}^{i}$. Then, by noting that $\langle\hat{x}^{i}\rangle=0$, since $\langle\hat{\psi}\rangle=0$, from Eq. \eqref{sec4_eq_position}, we obtain
\begin{eqnarray}\label{sec4_eq_position_dispersion}
\langle (\Delta\hat{x}^{(i)})^{2}\rangle_{\textrm{ren}} = \dfrac{q^{2}}{2m_{0}^{2}}\int_{0}^{\tau}dt\int_{0}^{\tau}dt'\int_{0}^{t}dt_{1}\int_{0}^{t'}dt_{2}g^{ij}_{1}g^{ij}_{2}\partial_{j_{1}}\partial_{j_{2}}G^{(1)}_{\textrm{ren}}(z_{1},z_{2}),
\end{eqnarray}
where the coincidence limit operation is implied. Eq. \eqref{sec4_eq_position_dispersion} is the dispersion in the coordinates of a point particle in the Einstein's universe, which are induced by quantum vacuum fluctuations of a massless scalar field. 
Similar to the previous subsection, we obtain that the dispersion in the vacuum state corresponds to the mean value of the coordinate squared: $\langle (\Delta\hat{x}^{(i)})^{2}\rangle_{\textrm{ren}}=\langle (\hat{x}^{(i)})^{2}\rangle_{\textrm{ren}}$. In both cases, that is, for the momentum \eqref{sec4momentum_disp_gen_relation} and coordinates \eqref{sec4_eq_position_dispersion}, this is a consequence of the linear dependence of $p^{i}$ and $x^{i}$ on the field $\psi(x)$, as we can see from Eqs. \eqref{sec4eq03} and \eqref{sec4_eq_position}.

For the angular coordinate $\chi$, from Eqs. \eqref{sec4_eq_position_dispersion}, \eqref{metricEinsteinUniverse} and  \eqref{sec4eq_kernel}, we obtain that 
\begin{eqnarray}
\langle (\Delta\hat{\chi})^{2}\rangle_{\textrm{ren}} = \dfrac{q^{2}}{m_{0}^{2}a_{0}^{4}}\int_{0}^{\tau}dt\int_{0}^{\tau}dt'\int_{0}^{t}dt_{1}\int_{0}^{t'}dt_{2}K_{\chi}(z,z'),
\end{eqnarray}
which after solving the respective operations, in the coincidence limit, give us 
\begin{eqnarray}\label{sec4_eq_chi_dispersion}
\langle (\Delta\hat{\chi})^{2}\rangle_{\textrm{ren}} = - \dfrac{\bar{q}^{2}}{6\pi^{2}}\mathcal{F}(\tau_{a}),
\end{eqnarray}
where we have defined the dimensionless charge parameter,
\begin{eqnarray}\label{sec4_eq_dimensionless_charge}
\bar{q}=\dfrac{q}{m_{0}a_{0}},
\end{eqnarray}
as well as the auxiliary function
\begin{eqnarray}\label{sec4_eq_auxiliary_function_chi_disp}
\mathcal{F}(r)=\dfrac{r^{2}}{24}+\dfrac{r}{2}\cot\left(\dfrac{r}{2}\right)-1-\dfrac{1}{2}\ln\left[\dfrac{\sin(r/2)}{(r/2)}\right]^{2} + \dfrac{1}{2}\int_{0}^{r}du u \ln\left[\dfrac{\sin(u/2)}{(u/2)}\right]^{2}.
\end{eqnarray}

For the other two coordinates, $\theta$ and $\phi$, using Eq. \eqref{sec4_eq_position_dispersion} with $i=\theta$ and $i=\phi$, in addition to \eqref{metricEinsteinUniverse} and \eqref{sec4eq_kernel}, we find that 
\begin{eqnarray}
\langle (\Delta\hat{\theta})^{2}\rangle_{\textrm{ren}} = \dfrac{q^{2}\csc^{4}(\chi)}{m_{0}^{2}a_{0}^{4}}\int_{0}^{\tau}dt\int_{0}^{\tau}dt'\int_{0}^{t}dt_{1}\int_{0}^{t'}dt_{2}K_{\theta}(z,z')
\end{eqnarray}
and
\begin{eqnarray}
 \langle (\Delta\hat{\phi})^{2}\rangle_{\textrm{ren}} = \dfrac{q^{2}\csc^{4}(\chi)\csc^{4}(\theta)}{m_{0}^{2}a_{0}^{4}}\int_{0}^{\tau}dt\int_{0}^{\tau}dt'\int_{0}^{t}dt_{1}\int_{0}^{t'}dt_{2}K_{\phi}(z,z'),
\end{eqnarray}
whose solutions are, respectively,
\begin{eqnarray}\label{sec4_eq_theta_dispersion}
\langle (\Delta\hat{\theta})^{2}\rangle_{\textrm{ren}}=\csc^{2}(\chi)\langle (\Delta\hat{\chi})^{2}\rangle_{\textrm{ren}}
\end{eqnarray}
and
\begin{eqnarray}\label{sec4_eq_phi_dispersion}
\langle (\Delta\hat{\phi})^{2}\rangle_{\textrm{ren}}=\csc^{2}(\chi)\csc^{2}(\theta)\langle (\Delta\hat{\chi})^{2}\rangle_{\textrm{ren}}.
\end{eqnarray}
Eqs. \eqref{sec4_eq_chi_dispersion}, \eqref{sec4_eq_theta_dispersion} and \eqref{sec4_eq_phi_dispersion} give us the dispersion in the vacuum state for the coordinates $\chi$, $\theta$ and $\phi$ in the Einstein's universe, respectively. 

Similar to the study of the momentum dispersion, we now must obtain the dispersion for the respective physical coordinates. Observing the line element \eqref{metricEinsteinUniverse}  we can verify that the physical distances or lengths, $\mathcal{z}_{i}$, are given by 
\begin{eqnarray}\label{sec4eq10}
d\mathcal{z}_{i} = \left\{d\mathcal{z}_{\chi} ; d\mathcal{z}_{\theta} ; d\mathcal{z}_{\phi}\right\} = \left\{a_{0}d\chi ; a_{0}\sin(\chi)d\theta ; a_{0}\sin(\chi)\sin(\theta)d\phi\right\},
\end{eqnarray}
which correspond to the modulus of the length elements in the geometric space defined by the Einstein's universe. Thus, from Eqs. \eqref{sec4eq10}, \eqref{sec4_eq_chi_dispersion}, \eqref{sec4_eq_theta_dispersion} and \eqref{sec4_eq_phi_dispersion}, we obtain that
\begin{eqnarray}\label{sec4eq11}
\langle (\Delta\hat{\mathcal{z}}_{i})^{2}\rangle_{\textrm{ren}}=a_{0}^{2}\langle (\Delta\hat{\chi})^{2}\rangle_{\textrm{ren}}.
\end{eqnarray} 
Also similar to the case of momentum dispersion, the result \eqref{sec4eq11} shows that the dispersions for the respective physical lengths are all equal. This fact exposes again the manifestation of homogeneity and isotropy properties of the Einstein's universe.

The temporal behavior of Eqs. \eqref{sec4_eq_chi_dispersion}, \eqref{sec4_eq_theta_dispersion} and \eqref{sec4_eq_phi_dispersion} correspond to the behavior of the function $\mathcal{F}(\tau_{a})$ in \eqref{sec4_eq_auxiliary_function_chi_disp}, up to  multiplicative constants. Here we note the presence of the same temporal divergences which occur in momentum dispersion, that is, for time values $\tau_{a}=2\pi n$, or in the dimensonal form $\tau=2\pi a_{0}n$. 
In addition, we also note that for $\tau_{a}\rightarrow 0$ as well as in the limit $a_{0}\rightarrow\infty$ we obtain that $\mathcal{F}(\tau_{a})=0$. Consequently, Eqs. \eqref{sec4_eq_chi_dispersion}, \eqref{sec4_eq_theta_dispersion}, \eqref{sec4_eq_phi_dispersion} and \eqref{sec4eq11} also vanish. As we know, the null result for $\tau_{a}=0$ is a consequence of the classical assumptions, which in this case corresponds to choosing $x^{i}(t)=0$ for $t=0$ in Eq. \eqref{sec4_eq_position}. On the other hand, in the case of the limit $a_{0}\rightarrow\infty$ it refers to the fact that there is no IQBM for the renormalized Minkowski vacuum.
The singular behavior of Eqs. \eqref{sec4_eq_theta_dispersion} and \eqref{sec4_eq_phi_dispersion} with respect to the angles $\chi$ and $\theta$ is possibly a consequence of the compact geometry of space.

In the analyzes of the momentum dispersions we have considered the hypothesis that temporal variations in the particle coordinates are negligible. Consequently, these simplifications will impose restrictions on our results, in other words, on the free parameters present in the expressions.  In order to obtain some insights about the small displacements condition in the present study, it is instructive to briefly recall some examples from the literature in which this condition arises.

In Ref. \cite{de2014quantum}, considering the Minkowski spacetime, the one-dimensional case of a point particle coupled to a massless scalar field in the presence of a point-like reflecting plane placed at $x=0$ was studied. There, the small displacements condition is interpreted mathematically as a restriction on the relative dispersion, $\left|\langle(\Delta x)^{2}\rangle_{\textrm{ren}} /x^{2}\right|\ll 1$, where $x$ is the distance of the particle from the point-like plane. A similar condition occur for the case of a point particle coupled to a massless escalar field in (3+1) dimensions confined by two parallel planes or by a one-dimensional compactification \cite{Ferreira:2023uxs}. In the case of the confinement via compactification process, the mathematical condition is such that $\left|\langle(\Delta x)^{2}\rangle_{\textrm{ren}} /d^{2}\right|\ll 1$, where $d$ is the compactification length. In both cases mentioned, in order to satisfy the approximation of neglecting the temporal variations of the coordinates, it is required that the relative (dimensionless) dispersion be smaller than unity.

\begin{figure}[h]
\includegraphics[scale=0.54]{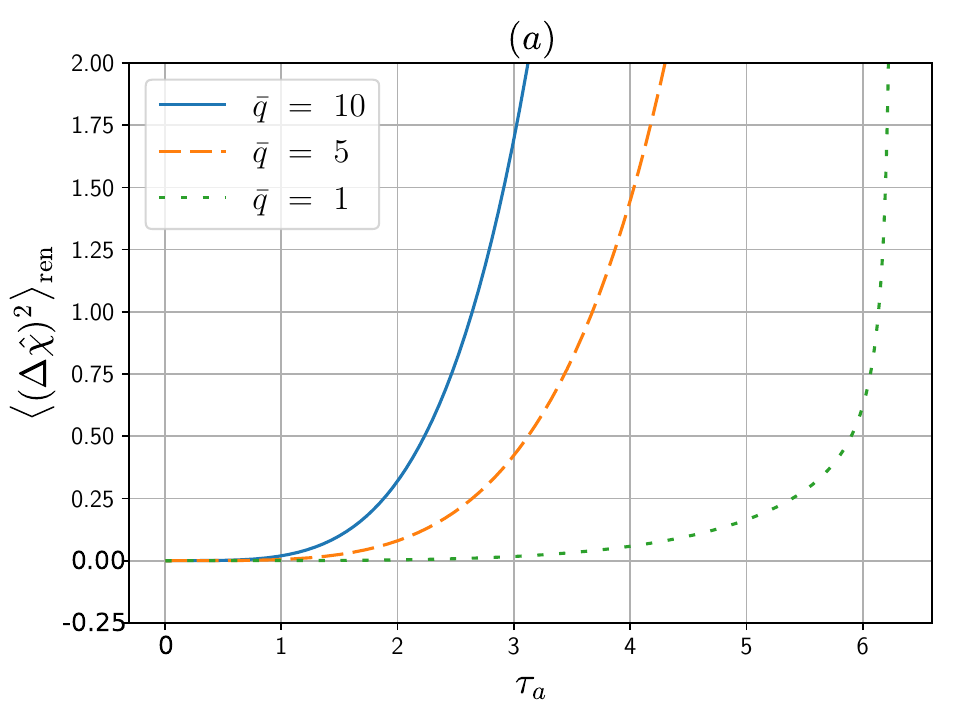}
\includegraphics[scale=0.54]{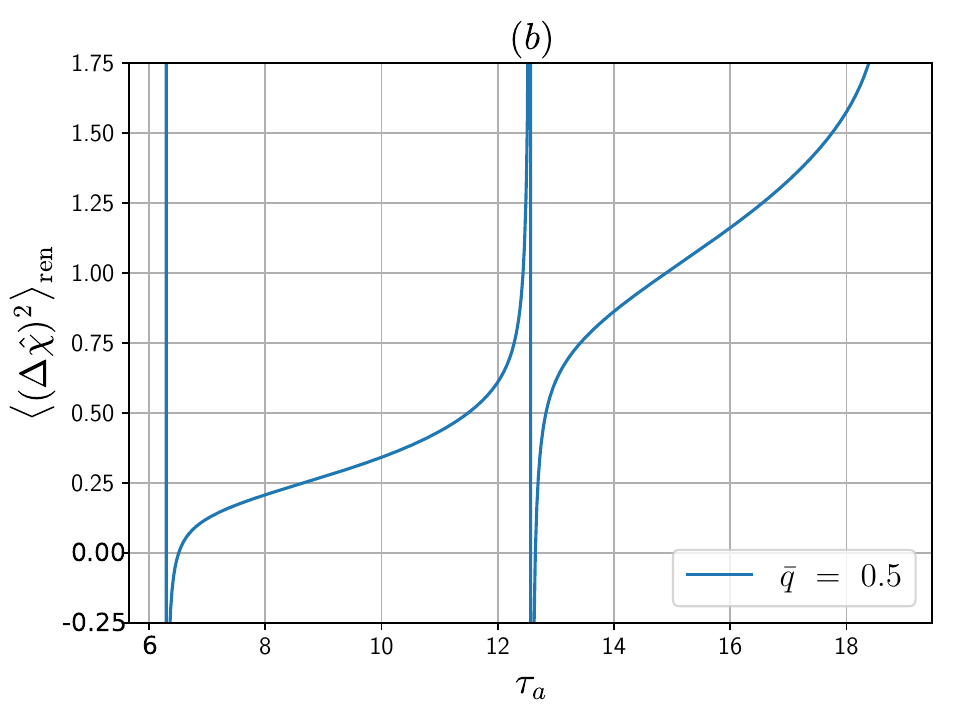}
\caption{Relative dispersion \eqref{sec4_eq_chi_dispersion} as a function of the dimensionless time $\tau_{a}$ for arbitrary values of the dimensionless charge: (a) $\bar{q}=10$, $\bar{q}=5$, $\bar{q}=1$ and (b) $\bar{q}=0.5$. }\label{fig_rel_disp}
\end{figure}

In practical terms the restriction on the relative dispersion in our case, from Eq. \eqref{sec4eq11}, is written as
\begin{equation}
\frac{\langle (\Delta\hat{\mathcal{z}}_{i})^{2}\rangle_{\textrm{ren}}}{a_0^2}=\langle (\Delta\hat{\chi})^{2}\rangle_{\textrm{ren}}\ll 1.
\label{SDC}
\end{equation}
Hence, the above expression represents the small displacement condition for our analysis. In Fig.\ref{fig_rel_disp} we have plotted, as a function of $\tau_a$, the relative dispersion above by making use of Eq. \eqref{sec4_eq_chi_dispersion} and observed the time value for which it is equal to unity. This time value will correspond to the upper bound value for which the condition is valid. Note that for the plots we assume distinct values for the parameter $\bar{q}$.

%
%
%
%Probably, a more accurate interpretation about this aspect, that is, angular divergences, can be provided by analyzing the IQBM for a massless scalar field confined by a spherical surface.

Based on this discussion, for each value of $\bar{q}$ chosen, we observe that the curves shown in the plots of Fig.\ref{fig_rel_disp} say that the condition on Eq. \eqref{sec4_eq_chi_dispersion} requires an upper bound on the dimensionless time parameter $\tau_{a}$ corresponding to $\langle(\Delta\hat{\chi})^{2}\rangle_{\textrm{ren}}=1$. In Table \ref{tableSDC} it possible to see the upper bound values for $\tau_{a}$ in the cases exhibited in Fig.\ref{fig_rel_disp}a. These graphs have been considered in the range $0<\tau_{a}<2\pi$. In other words, the values shown in Table \ref{tableSDC} represent the upper bound value for the time $\tau_a$ for each chosen $\bar{q}$. For instance, in the case $\bar{q}=10$, this occurs for $\tau_a =2.642$. Therefore, the time interval considered in our analyses should be such that $0<\tau_a <2.642$, as within this interval the condition of small displacements, Eq. \eqref{SDC}, is satisfied. If we consider the value $\bar{q}=5$, compared to the previous case, we observe that the upper time limit increases to $0<\tau_a <3.675$, consequently indicating that the validity range of condition \eqref{SDC} also increases. The same situation occurs for $\bar{q}=1$. Thus, these results reveal that the smaller the values of $\bar{q}$ chosen the better the effectiveness of our analysis, since the set of values for $\tau_{a}$ in the range taken into consideration increases while at the same time does not violate the small displacement condition in Eq. \eqref{SDC}.
A conclusion similar to this one was reached in Refs. \cite{de2014quantum} and \cite{Ferreira:2023uxs} for a point particle coupled to a massless scalar field in Minkowski spacetime.
%
%We conclude that the margin of applicability of our results increases as $\bar{q}$ decreases.
%
%
%
%
%
%
\begin{table}[h]
\centering
\caption{Approximate time values $\tau_{a}$ for the upper bound limit of the condition $\langle(\Delta\hat{\chi})^{2}\rangle_{\textrm{ren}}\ll 1$.}\label{tableSDC}
	\begin{tabular}{ccc}
	\toprule
	\hspace{0.5cm}$\bar{g}$\hspace{0.5cm} &\hspace{2cm}& \hspace{0.5cm}$\tau_{a}$\hspace{0.5cm} \\ \midrule\midrule
	 		10 							  & 		   & 2.642\\ \midrule
	    	5							  & 		   & 3.675\\ \midrule
	  	    1							  & 		   & 6.135\\
	\bottomrule\bottomrule
	\end{tabular}
\end{table}

Although our analysis in the plots above for the relative dispersion $\langle(\Delta\hat{\chi})^{2}\rangle_{\textrm{ren}}$ has been restricted to the interval $0<\tau_{a}<2\pi$, it can also been extended to subsequent intervals, such as the one shown in Fig.\ref{fig_rel_disp}b where $2\pi<\tau_{a}<6\pi$. We can see that, by taking $\bar{q}=0.5$, in the interval $2\pi<\tau_{a}<4\pi$ our investigation is still effective. However, in the interval $4\pi<\tau_{a}<6\pi$, the effectiveness of our analysis is reduced since the condition \eqref{SDC} is only satisfied for values of the dimensionless time up to $\tau_{a}\simeq 15$. Note that the vertical blue lines in the plot of Fig.\ref{fig_rel_disp}b indicates round trip-like divergencies.

To end this section, we would like to comment on the connection between the results from the present subsection and the ones from Section \ref{Sec4_momentum_dispertions}. The discussions above indicate that the validity and effectiveness of the analysis of the momentum dispersion results will be limited to the intervals of $\tau_{a}$ (Table \ref{tableSDC}) that do not violate the condition \eqref{SDC}. For example, in the case of $\bar{q}=10$, we have $\tau_{a}\approx 2.64$, so the analysis of $\langle(\Delta \hat{\mathcal{p}}^{i})^{2}\rangle_{\textrm{ren}}$ are limited to this time interval, because otherwise, if $\tau_{a}>2.64$, the condition of small displacements would be violated, leading to a contradiction, as this is an assumption upon which we have developed our studies. Thus, in order for the graph in Fig.\ref{figMomentumDispersion} not violate the small displacements condition we would have to take $\bar{q}\simeq 0.24$.

\section{Conclusions and final remarkers}

In this work, by assuming the small displacement condition, we have investigated the IQBM of a point particle coupled to a massless scalar field in a curved spacetime, in which the background geometry has a closed curvature and represents a static Universe. It is in fact the homogeneous and isotropic FRW Universe, with a constant scale factor, and it is commonly known as the Einstein's universe.
As a consequence of the homogeneity and isotropy of the spacetime we have obtained that all nonzero momentum dispersion components are equal, a result that also occurs for the physical position components -- see Eqs. \eqref{sec4eq09} and \eqref{sec4eq11}.

We have also shown that the expressions for the dispersion in the momentum and position of the point particle present round trip-like divergencies when $\tau=(2\pi a_{0})n$ ($n=1,2,3,...$), which can be seen from Figs.\ref{figMomentumDispersion} and \ref{fig_rel_disp}, in addition to Eqs. \eqref{sec4eq09}, \eqref{sec4eq_mdisp_chi_F}, \eqref{sec4_eq_chi_dispersion} and \eqref{sec4_eq_auxiliary_function_chi_disp}. An interesting aspect of the dispersion in the momentum is that it is positive and increases more and more with the time interval $2\pi a_{0}$ without distorting the shape of its curve. This nontrivial behavior can be seen in Fig.\ref{figMomentumDispersion}.
These divergences are consequence of the compact or closed topology of the Einstein's universe, which causes an effect similar to that of compactification as analyzed by the authors in Ref. \cite{Ferreira:2023uxs}.

As to the dispersion in the position components of the point particle that undergo quantum brownian motion, we have analyzed in what conditions the small displacement condition is effective and have shown that the dimensionless charge parameter $\bar{q}$ plays a crucial role in the investigation. In other words, as we take small values for $\bar{q}$ the values the dimensionless time $\tau_a$ can assume increases, in the interval $0<\tau_{a}<2\pi$. Essentially, $\bar{q}=1$ is enough to have this whole interval covered, as we can see in Fig.\ref{fig_rel_disp}a.  We have also shown that by extending the values of $\tau_a$ beyond the interval $0<\tau_{a}<2\pi$, the effectiveness of our analysis tends to be reduced, as it can be seen in Fig.\ref{fig_rel_disp}b. In this plot, as we have pointed out, the vertical blue lines represent round trip-like divergencies.

{\acknowledgments}

E.J.B.F would like to thank the Brazilian agency Coordination for the Improvement of Higher Education Personnel (CAPES) for financial support. H.F.S.M is partially supported by the National Council for Scientific and Technological Development (CNPq) under grant No 311031/2020-0.

%\bibliographystyle{JHEP}
%\bibliography{refA3.bib}

\end{document}